\documentclass[sn-nature]{sn-jnl} %{article}
\usepackage{graphicx}%
\usepackage{multirow}%
\usepackage{amsmath,amssymb,amsfonts}%
\usepackage{amsthm}%
\usepackage{mathrsfs}%
\usepackage[title]{appendix}%
\usepackage{xcolor}%
\usepackage{textcomp}%
\usepackage{manyfoot}%
\usepackage{booktabs}%
\usepackage{algorithm}%
\usepackage{algorithmicx}%
\usepackage{algpseudocode}%
\usepackage{listings}%

\newcommand{\eg}{\`e }

\begin{document}

\title{{\bf Cosmological implications of the minimum viscosity principle 
}}

\author[1]{\fnm{P. G.} \sur{Tello}}
\author[2]{\fnm{Sauro} \sur{Succi}}
\email{sauro.succi@iit.it}

\affil[1] {CERN, Geneva, Switzerland}

\affil*[2]{\orgdiv{Center for Life Nano- \& Neuro-Science}, \orgname{Italian Institute of Technology (IIT)}, \orgaddress{\street{viale Regina Elena 295}, \city{Rome}, \postcode{00161}, \country{Italy}}}

%\author{P.G. Tello$^{1}$, S. Succi$^{2}$\\
%(1) CERN, Geneva, CH\\
%(2) IIT, Roma, Italy,\\
%}

%\begin{abstract}

\abstract{It is shown that black holes in a quark gluon plasma (QGP) obeying
minimum viscosity bounds, exhibit a Schwarzschild radius in close match 
with the range of the strong force. 
For such black holes, an evaporation time of about $10^{16}$ secs is estimated,
indicating that they would survive by far the quark-gluon plasma era, namely
between $10^{-10}$ and $10^{-6}$ seconds after the big bang.  
On the assumption that the big-bang generated 
unequal amounts of quark and antiquarks, this suggests
that such unbalance might have survived to this day 
in the form of excess antiquark nuggets hidden to all but
gravitational interactions.
A connection with the saturon picture, whereby minimum viscosity 
regimes would associate with the onset coherent quantum field
structures with maximum storage properties, is also 
established, along with potential implicationd for quantum computing
of classical sytems.}

%\end{abstract}

\maketitle

\section{Introduction}

The predominant model describing the origin of our Universe posits that 
approximately fourteen billion years ago, all matter emerged from a 
triggering event known as the Big Bang. 
After its occurrence, a searing quark-gluon plasma (QGP) materialized, giving rise 
to the fundamental constituents of matter: baryons (comprising protons and neutrons forming 
atomic nuclei), electrons, and photons \cite{1,2}. 

The existence of the QGP phase, where traditional particles dissolve into their 
fundamental constituents, has been experimentally demonstrated through 
relativistic heavy ion collisions [3,4]. As the plasma cooled down, the formation of 
light atoms ensued, while the synthesis of heavier elements took place later, within 
the cores of stars. However, a significant challenge arises in comprehending how a 
universe predominantly composed of baryons emerged from pure energy. 
Common reactions generating baryons typically yield corresponding anti-baryons, presenting 
a riddle regarding the apparent scarcity of antimatter in our observable universe. 
An interesting hypothesis suggests a baryon-number-violating reaction during the early 
stages, resulting in the observed matter-antimatter asymmetry. 
It proposes that alongside the formation of baryons, the Big Bang generated quark 
nuggets—both matter and antimatter varieties—in {\it unequal quantities} \cite{5,6,7}. 
Quark nuggets house substantial amounts of nuclear matter, potentially 
contributing to a significant baryonic number. 

In this scenario, the overall count of baryons minus antibaryons in the Universe 
remains at zero, with ordinary matter predominantly consisting of baryons, while 
the excess antibaryons reside within quark nuggets. 
Also, besides their intriguing potential as dark matter candidates, quark 
nuggets have been hypothesized as sources of black holes in the early 
Universe \cite{8,9,10,11}. 

This Letter delves into speculation about the potential 
formation of black holes within the quark-gluon plasma, aiming 
to contribute to the ongoing discussion, particularly 
from the perspective of the minimum viscosity principle.

\section{QGP minimum viscosity}

The quark-gluon plasma (QGP) is a strongly-interacting 
quantum relativistic fluid
which has been shown to saturate the minimum 
viscosity bound (MVB):
\begin{equation}
\label{MVB}
\frac{\eta}{s} \ge \frac{1}{4 \pi} \frac{\hbar}{k_B}
\end{equation}
where $\eta$ is the dynamic viscosity and $s$ the entropy per unit volume 
Rearranging in terms of the kinematic viscosity, 
$\nu=\mu/\rho$, $\rho$ being the QGP density, this 
reads as follows:
\begin{equation}
\label{MVB2}
\nu \ge \nu_{MVB} = \sigma \frac{\hbar}{m} 
\end{equation}
where the entropic coefficient $\sigma = \frac{log W}{4 \pi}$
follows from the Boltzmann's relation $S = k_B log W$.

The ultimate meaning of the MVB is that the mean free path of 
the quantum excitations cannot exceed the De Broglie 
wavelength, times the entropic factor $\sigma$, namely:
\begin{equation}
\lambda_{mfp} \ge \sigma \lambda_B
\end{equation}
  
The presence of the entropic term $\sigma$ reflects the gravitational
roots of the MVB, as originally exposed by the celebrated
duality between gravity and conformal field theory \cite{MALDA}.
Based on such duality, the kinematic viscosity of the QGP 
can also be cast in the form:
\begin{equation}
\label{NUQ}
\nu \sim  \frac{\eta}{Ts} c^2
\end{equation}
where $c$ is the speed of light and $T$ is the QGP temperature.

Based on the aforementioned duality, the kinematic viscosity of a QGP
confined in a region of size $R$ can also be expressed
in terms of the gravitational constant as follows:
\begin{equation}
\label{NUG}
\nu \sim \frac{c^3}{GR}
\end{equation}
By taking $s=S/R^3$ and using the Bekenstein entropy bound $S \sim k_B (R/L_p)^2$,
$L_p = (G\hbar/c^3)^{1/2}$ being the Planck length, one readily obtains  
\begin{equation}
\label{NUP}
\nu \sim c^2 \tau_q
\end{equation}
where $\tau_q = \hbar/k_b T$ is the quantum thermal relaxation time.
The latter shows that the QGP abides by the so-called {\it Planckian transport}
mechanism, whereby the resistivity scales linearly with the temperature.  
Planckian transport makes the current object of intensive investigation in condensed matter
since several exotic forms of electronic transport, including high-Tc superconductors,
seem to share into this intriguing regime \cite{20}.

\section{QGP and black holes}

Let us assume the that primordial QGP gives rise to a black hole
of mass $M$ and Schwartzschild $R_s = GM/c^2$. 
By computing the mass as $M = (\mu/\nu) R^3$, eq. (\ref{NUQ}) 
we obtain $R_s \sim GR^3 sT/c^4$. 
Using Bekenstein's bound again delivers
$R_s \sim c \tau_q$, showing that the Schwartzschild radius 
of the black hole is basically the mean free path of the QGP excitations 
saturating the Bekenstein bound. 
Differently restated, in view of the relation 
(\ref{NUP}), we also have: 
\begin{equation}
R_s \sim \nu/c
\end{equation}
By taking the minimal value $\nu \sim 10^{-7}$ ($m^2/s$), we obtain 
$R_s \sim 10^{-15}$ ($m$), which compares tightly with the size of 
the proton, namely the range of strong interactions.

Incidentally, by treating the QGP as a fluid, the associated Reynolds 
number is given by $$\mathcal{R} = c R_s/\nu = 1,$$ indicating that
the size of the black-hole corresponds to the Kolmogorov length
of a hypothetical (inverse) turbulent enstrophy cascade, 
initiated by the gravitational collapse of the QGP \cite{21,22}.
As an aside, it is interesting to observe that the condition for 
the survival of coherent structures at a generic scale $l$, $Re(l)>1$, can 
be interpreted as an "uncertainty" relation, namely
$v(l) l \ge \nu$. This means that in order to survive dissipation, a
turbulent eddy of size $l$ must feature a velocity fluctuation above 
a threshold $\bar v(l) = \nu/l$.
Clearly, only singular fluctuations $v(l) \sim l^{\alpha}$, $\alpha<-1$ 
can meet this constraint down to the UV limit $l \to 0$.
In actual facts, turbulent fluctuations feature positive scaling exponents,
that is $\alpha = 1/3$ and $\alpha=1$ in three and two dimensions, respectively
\cite{23,VOLO} 

% ----------------------------------
\begin{figure}
\centering
\includegraphics[scale=0.3]{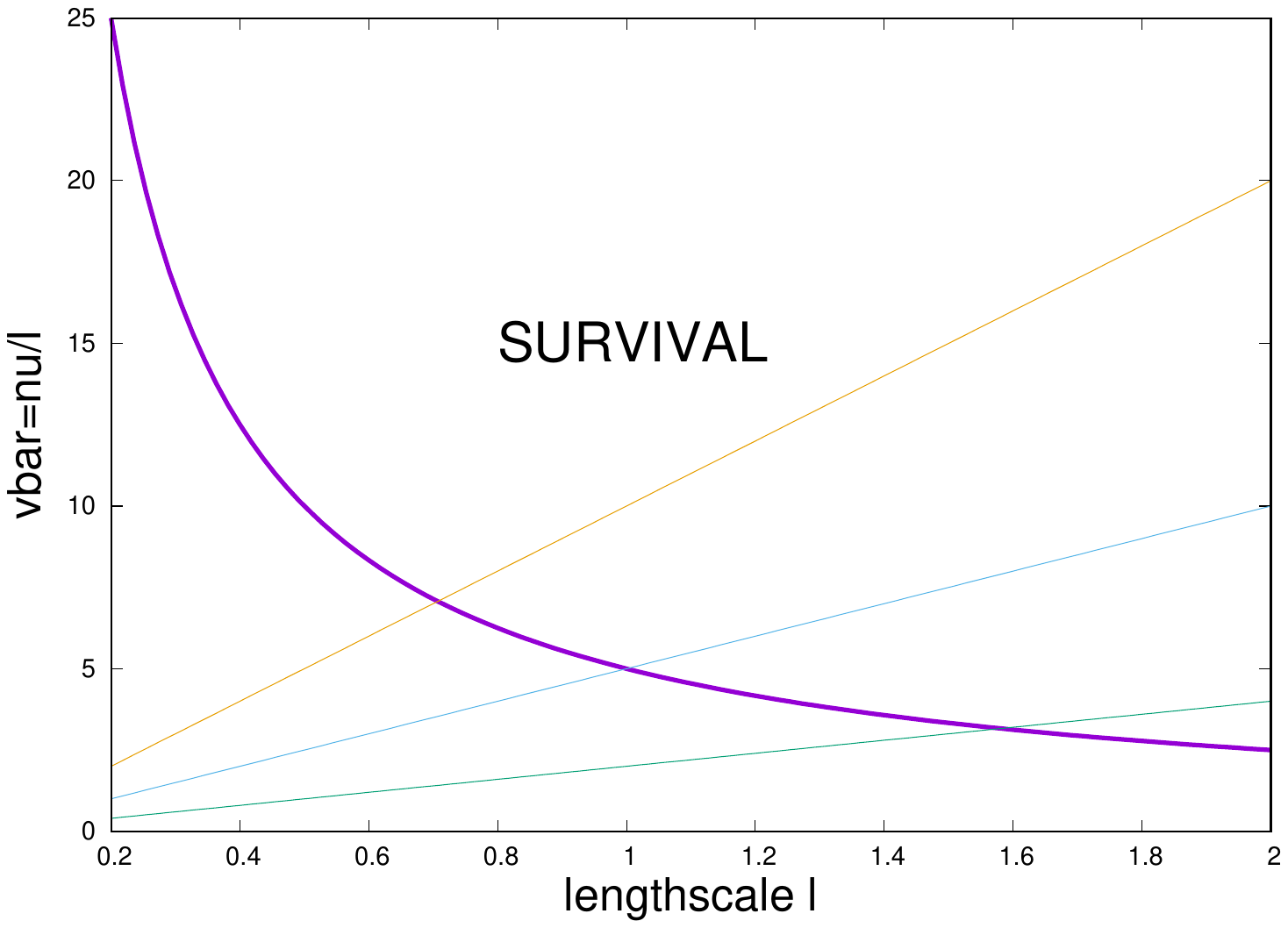}
\caption{The survival region of coherent structures in 2d turbulence.
Only eddies which pass the bar $\bar v (l) = \nu/l$ manage to survive.
The straight lines indicate 2d turbulent eddies with $v(l) = v(L)l/L$
where $L$ is the infrared scale of the domain.
The crossover $v(l_d)=\bar v(l_d)$, marks the dissipative scale $l_d$.
Eddies with $l<l_d$ do not survive dissipation and from a hydrodynamic
standpoint, their information content is irreversibly lost. 
}
\end{figure}

Hence there is always a finite scale, the Kolmogorov dissipative
length $l_d$, below which survival of coherent structures is no longer possible.
In classical physics it is possible, at least in principle, to
send the kinematic viscosity to zero (infinite Reynolds number limit), in which
case the dissipative scale also goes to zero.
More precisely, for a turbulent flow in a region of global size $L$,
one has $l_d = L/\mathcal{R}^{\beta}$, with $\beta=1/2$ 
in two dimensions and $\beta=3/4$ in three.

However, the minimum viscosity principle forbids the infinite Reynolds number limit.

Indeed, discounting entropic contributions and
setting $\nu \sim \hbar/m$, one recovers exactly the 
Heisenberg relation $\delta p \delta l \ge \hbar$.
This shows that the minimum viscosity picture provides a formal
bridge between turbulent and quantum fluctuations.
Of course this analogy must be taken with a huge pinch of salt, for the
physics of quantum and turbulent fluctuations are pretty distinct from each other.
Yet, the formal analogy might hint at a unifying thread in terms 
of tyhe correspoding information loss mechanisms. 
Just like no information can be gleaned from a quantum system in a
phase-space box of size below $\hbar$, coherent information on
a turbulent flow is lost on a phase-space box of area below $\nu$. 

Next, let us consider the evaporation time for such a QGP-BH, namely
\begin{equation}
t_{ev} \sim (\frac{T_p}{T})^2 \; \tau_q
\end{equation}
By taking $k_BT \sim 100$ MeV and recalling that $T_p \sim 10^{32}$ K, this returns
$T_{ev} \sim 10^{16}$ seconds, showing that such minimum 
viscosity BHs would long survive the QCD era in the history of the Universe.
Hence they could still be with us and serve as segregating units 
for dark anti-baryonic matter, although this is a mere speculation at this stage.

\section{Turbulence-driven black-hole formation}

The formation of BH out of density fluctuations requires 
a strong-fluctuation regime:
\begin{equation}
\label{SF}
\frac{\delta \rho}{\rho} \sim 1
\end{equation}
It is therefore of interest to inspect under what conditions would 
such a regime possibly occur.
The parameter controlling the amplitude of density fluctuations is the Mach number
$Ma = u/c_s$, where $u$ is the macroscopic flow velocity and $c_s$ is the sound speed.
In particular, strong fluctuations fulfilling (\ref{SF}) require supersonic flows, $Ma \ge 1$.
Indeed such type of fluctuations have been observed in astrophysical plasmas and 
simulations of compressible MHD turbulence alike \cite{}.
Of course, this does not mean that the same is true for general relativity, but since
it is known that BH horizons display turbulent regimes, this cannot be ruled out either.

To make this conjecture a bit more quantitative, let us consider the Universe at the 
QCD epoch, $t=10^{-6}$ seconds, with an estimated radius $R=ct \sim 10^2$ meters.
On the assumption of a radiative equation of state $c_s = c/\sqrt{3}$ and a sonic flow
with $u \sim c_s$, the corresponding Reynolds number is estimated as
$Re \sim 10^8 \times 10^2 /10^{-7} = 10^{17}$. 
The associated dissipative Kolmogorov scale is $l_d = R/Re^{3/4} \sim 10^{-11}$ meters, four 
orders of magnitude above the Schwarzschild radius
of the putative QGP black holes. As a result, we conclude that
the density fluctuations triggering the BH formation are 
not turbulent but rather "molecular" in character.
The mean free path of such molecular fluctuations is readily estimated 
as $\lambda_{mfp} = Ma \frac{R}{Re}$, namely $\sim 10^{-15}$ meters, which is exactly
the Schwarzschild radius computed in the previous section.
To be noted that the condition $Ma \sim 1$ is still required.

The picture emerging from this analysis is that of an expanding Universe 
which at the QCD epoch can be regarded as a compressible turbulent flow with Reynolds number
$Re \sim 10^{17}$, feeding coherent structures down to the scale 
of $10^{-11}$ meters, four orders of magnitude above the QCD scale.
The mean free path of the "molecular gas" below the Kolmogorov scale is however
in close match with the QCD scale, indicating that QGP black-holes could be
triggered but strong density fluctuations of the QGP.
The density fluctuations of the quark gluon plasma can be estimated 
following the procedure described in \cite{POL}. 
Averaging the Fermi-Dirac (quarks) and Bose-Einstein (gluons) distributions 
with a gaussian filter 
of size $\Delta x^3 \Delta t$, delivers the following expression
\begin{equation}
\label{FLUCT}
\delta \equiv \frac{\langle \delta n^2 \rangle}{\langle n \rangle^2} = 
\frac{1}{\langle n \rangle}
\frac{1}{(2 \pi)^{3/2}} 
\frac{1}{\Delta x^2 \sqrt{\Delta x^2 + c^2 \Delta t ^2}} 
\end{equation}
By taking $\langle n \rangle = 1$ ($fm^{-3})$, and $\Delta x = c \Delta t =1$ ($fm$), we obtain
$\delta \sim 0.05$. Hence, $\Delta x \sim 0.3$ delivers $\delta \sim 1$   
The above formula assumes massless excitations propagating at 
luminal speed $v(p)=c$. Based on the above formula, massive particles with 
$v(p)<c$ would lead to slightly larger density fluctuations.
Of course this is not a proof but just a plausibilty scenario, and 
yet, one which appears to be consistent not only in principle but 
also in terms of the numerical values of the main observables in point.

This said, alternative scenarios are available, as we are going to
discuss in the next section.

\section{The QGP-BH-Saturon connection}

Recent arguments suggest that quantum field structures, known as {\it saturons}, 
possessiing maximal microstate entropy within the constraints of unitarity, might bear 
striking similarities to black holes \cite{14,15,16,17,18,19}, in that 
their entropy scaling mirrors that of the Bekenstein-Hawking formula. 
Furthermore, saturons undergo decay in accordance with Hawking's 
thermal radiation, with a decay rate proportional to the inverse of their size. 
The underlying concept posits that field-theoretic entities endowed 
with maximal information storage capacity, exhibit universal 
characteristics, regardless of their specific microscale structure. 
These characteristics are naturally formulated in the language of 
Goldstone modes, thereby establishing a direct connection
between saturons and symmetry breaking (e.g. Poincar\eg symmetry). 
In the specific case of BHs, the saturon is interpreted 
as a Bose-Einstein condensate of soft gravitons of wavelength $R$.

In the following, we merely put together the basic 
facts of the saturon picture and show that the aforementioned QGP
black-holes match the requirements of saturon's theory.   

The maximum information capacity of a saturon of 
radius $R$ is given by:
\begin{equation}
S_{max} \sim f^2 R^2
\end{equation}
where $f \sim \sqrt{N/R}$ is the decay constant of the 
Goldstone saturon, $N$ being the number of true-vacua of the
broken $SU(N)$ symmetry. Each of these true-vacua corresponds to a
microscopic realization of the maximum-entropy macroscopic state. 
In the unitary limit, the entropy saturation imposes the 
saturon coupling strength scales inversely with $N$, namely:

\begin{equation}
\label{UNI}
\alpha \sim 1/N
\end{equation}

It is now readily checked that the minimum viscosity QGP black-hole discussed
in the previous sections does indeed obey the unitarity limit (\ref{UNI}).
To this purpose, let us write $N \sim f^2 R^2$, and recall that $f \sim 1/G_{Gold}$,
$G_{Gold}$ being the Goldstone coupling.
By equating $G_{Gold} = G$, the gravitational constant, and expressing 
$R^2$ via the MVB relation
$(\nu/c)^2$, one can readily check that the saturon coupling 
strength $\alpha \sim G/R^2$ obeys indeed the unitary relation (\ref{UNI}).
This shows that a BH formed by a minimum viscosity QGP does indeed fit the
requirements of the saturon picture, thereby corroborating the portrait of
such a BH as a Bose-Einstein condensate of $N$ soft gravitons.

%The maximum storage property of saturons naturally invites the idea that they
%might provide appealing hardware platforms for quantum computing.

\subsection{Connections to quantum computing of classical systems}

We would like to close this paper with a few considerations regarding the possible
role of saturons as potential devices for the quantum simulation 
of nonlinear classical systems.
In his epoch-making 1982 paper \cite{FEY}, Feynman famously proclaimed that
{\it "Nature isn't classical, and if you want to make a simulation of nature you'd better
make it quantum mechanical...and it is a beautiful problem because it doesn't look so easy"}.
Feynman only implicitly hinted, in the final part of his famous sentence, at the
fact that while it is true that Nature  isn't classical, it is equally true that 
it has a very strong and built-in drive to become such at sufficiently 
large scales and energies.
In fact, this innate drive towards classicalization is precisely
the main hurdle towards the viable realization of quantum computers 
beyond the realm of theorems and complexity estimates.

A further question that Feynman apparently didn't address is whether quantum computers
can show any advantage in solving {\it classical} problems as well, turbulence and 
general gravitation being two outstanding examples in point.
Such question has only recently been recently tackled by the quantum 
computing community \cite{CHILDS,PALMER}.
This is a formidable challenge on top of a formidable challenge, since besides 
the well-known issues of decoherence and noise, the quantum simulation of 
classical fluids faces with two additional fundamental issues not shared by 
quantum systems: nonlinearity and dissipation \cite{EPL_QC}.

Several strategies have been developed in the recent years to handle
both non-linearity and dissipation, but for the time being, none of them 
has led to a practically viable quantum algorithm.
This is due to a number of reasons, a prominent one being that 
present-day quantum hardware is based on genuinely quantum systems
designed to withstand dissipation instead of embracing it.

Saturons offer maximum storage and information retrieval
capacity, but whether they can also process quantum
information in a way consistent with the requirements of the quantum 
simulation of nonlinear dissipative systems, remains a completely
open question at this point.
Since they are quantum objects compatible with the minimum 
viscosity principle, one may hope that they could  
indeed support non-linear and non-unitary qubit operation
out of reach to quantum computers based on genuinely quantum physical systems.

%The odds are fully open, including the possibility of "classical advantage" namely 
%that no quantum computer can ever beat a classical one in computing 
%nonlinear classical field theories.
%While definitely unwelcome from the practical point of view, such "classical advantage"
%might have interesting implications for fundamental science. 
%Emergent physics, life in the first place, vitally depends on nonlinearity. 
%Hence, should nonlinearity prove too hard to be quantum-computable,
%the emergence of classical physics could then be interpreted as a necessary
%built-in feature of quantum mechanics. A similar argument might then apply to the 
%alleged failure of gravity to become quantum, a perspective which is gathering 
%increasing attention in the quantum-gravity community.

\section{Conclusions}

Summarizing, starting from the principle of minimum viscosity, we have
explored the possibility of black hole formation 
froma gravitational collapse of a quark gluon plasma (QGP).
Based on purely dimensional arguments, it is shown that such QGP would
obey Planckian transport, i.e. its resistivity scales linearly with
the temperature, similarly to other exotic states of quantum non-equilibrium
condensed matter systems, such as high-Tc superconductors.
It is also found that in such Planckian-transport regime, the
Schwarzschild radius of a QGP black-hole sits tightly within the
range of the gluon-mediated colour force between quarks. 
Under such conditions the estimated evaporation time 
is around $10^{16}$ seconds, indicating survival far beyond the QCD era in the
history of the Universe, i.e. between $10^{-10}$ to $10^{-6}$ seconds. 
It was also shown that the large density fluctuations required to initiate black
hole formation are compatible with a kinetic theory descriotion of
the QGP at the scale of a fraction of femtometer.
Finally, always based form the minimum viscosity principle, supplemented by
purely dimensional arguments, it is shown that 
QGP driven black-holes  meet the prescriptions of the saturon 
hypothesis, namely they realize a maximum quantum information storage "device".
Possible implications for the quantum simulation of classical systems are briefly
discussed.

\section{Acknowledgements} 

SS wishes to acknowledge the hospitality and the highly stimulating 
environment of the Institut des Hautes Etudes Scientifiques, which
generously supported his stay through the funds of the 
2021 Balzan Prize for Gravitation: Physical and Astrophysical 
Aspects, awarded to T. Damour. 
He also wishes to acknowledge illuminating discussions with Prof. T. Damour.
Both authors wish to thanks Prof G. Dvali and K. Trachenko for many interesting
remarks.


\begin{thebibliography}{99}

\bibitem{1} K. Yagi, T. Hatsuda and Y. Miake, 
“Quark-Gluon Plasma: From Big Bang to Little Bang”, 
Cambridge University Press, 2008.

\bibitem{2} A. Riotto, 
“Baryogenesis ad Leptogenesis”, 
Journal of Physics: Conference Series 335 (2011) 012008.

\bibitem{3} S. Sarkar, H. Satz and B. Sinha, 
“The Physics of the Quark-Gluon Plasma: Introductory Lectures”, 
Springer 2010.

\bibitem{4} B. Sinha, S. Pal and S. Raha, 
“Quark-Gluon Plasma”, Springer-Verlag, 1990.

\bibitem{5} Y. Bai, A. J. Long and S. Lu, 
“Dark Quark Nuggets”, Phys. Rev. D. 99 (2019) 055047.

\bibitem{6} A. B. H. Bhattacharyya, S. Banerjee, S. K. Ghosh, S. Raha, B. Sinha and H. Toki, 
“Quantum chromodynamics phase transition in the early Universe and quark nuggets”, 
Pramana – J. Phys. 60, (2003) 909–919.

\bibitem{7} B. Layek, A. P. Mishra, A. M. Srivastava, and V. K. Tiwari, 
“Baryon inhomogeneity generation in the quark-gluon plasma phase”, 
Phys. Rev. D. 73, (2006) 103514.

\bibitem{8}
A. Atreya, A. Sarkar, and A. M. Srivastava, 
“Reviving quark nuggets as a candidate for dark matter”, 
Phys. Rev. D. 90 (2014) 045010.

\bibitem{9} Jan-e Alam, S. Raha, and B. Sinha, 
“Quark Nuggets as Baryonic Dark Matter”, 
The Astrophysical Journal, 513 (1999) 572.

\bibitem{10} X.Y. Lai and R.X. Xu, 
“Formation of the seed black holes: a role of quark nuggets?”, 
J. of Cosmology and Astroparticle Physics, 05 (2010) 028.

\bibitem{11} B. Sinha, “Hawking Radiation from Relics of the QCD Phase Transition—Strange Quark Nuggets, 
Primordial Black Holes, and White Holes”, 
Phys. Part. Nuclei 53 (2022) 159–166.

\bibitem{12} K. Trachenko, V. Brazhkin and M. Baggioli, 
“Similarity between the kinematic viscosity of quark-gluon plasma and liquids 
at the viscosity minimum”, SciPost Phys. 10 (2021) 5, 118.

\bibitem{13} G. Policastro, D. T. Son and A. O. Starinets, 
“Shear viscosity of strongly coupled N = 4 supersymmetric Yang-Mills plasma”, 
Phys. Rev. Lett. 87, (2001) 081601.

\bibitem{14} G. Dvali, O. Kaikov, and J. S. Valbuena Bermúdez, 
“How special are black holes? Correspondence with objects saturating unitarity 
bounds in generic theories”, Phys. Rev. D 105, (2022) 056013.

\bibitem{15} G. Dvali and O. Sakhelashvili, 
“Black-hole-like saturons in Gross-Neveu”, 
Phys. Rev. D. 105, (2022) 065014.

\bibitem{16} G. Dvali, 
“Entropy bound and unitarity of scattering amplitudes”, 
J. High Energy Phys. 03 (2021) 126.

\bibitem{17} G. Dvali, 
“Bounds on quantum information storage and retrieval”, 
Philos. Trans. A 380, (2021) 20210071.

\bibitem{18} G. Dvali, 
“Unitarity entropy bound: Solitons and instantons”, 
Fortschr. Phys. 69, (2021) 2000091.

\bibitem{19} G. Dvali, 
“Area law saturation of entropy bound from perturbative unitarity 
in renormalizable theories”, Fortschr. Phys. 69, (2021) 2000090.

\bibitem{MALDA} J. Maldacena,
The large-N limit of superconformal field theories and supergravity
Int. Journal of Theoretical Physics 38 (4), 1113-1133 (1999)
`
\bibitem{20} H Yang, A Zimmerman, L Lehner,
Turbulent black holes, 
Phys Rev Lett 114 (8), 081101, 83 (2015)

\bibitem{21} D Bini, S Kauffman, S Succi, PG Tello,
First post-Minkowskian approach to turbulent gravity,
Physical Review D 106 (10), 104007 2,(2022)

\bibitem{22} J. Zaanen,
Planckian dissipation, minimal viscosity and the transport in cuprate strange metals,
SciPost 6, 061 (2019)

\bibitem{23} U. Frisch, 
Turbulence, the legacy of A.N. Kolmogorov,
Cambridge Univ. Press, 1995

\bibitem{VOLO} G. Volovik,
Quantum Turbulence and Planckian dissipation,
JETP Letters, 115, 8, 461 (2022)
%ISSN 0021-3640, JETP Letters, 115, 8, 461 (2022)

\bibitem{POL} S. Mrowczynski,
Density fluctuations in the quark-gluon plasma,
Phys Rev C, 1518, (1998)

\bibitem{FEY} R. Feynman,
Simulating Physics with Computers,
Int. J. Mod Phys, 6, 467, (1982)

\bibitem{CHILDS} J.P. Liu, H. O. Kolden, H. K. Krovi et al,
Efficient quantum algorithm for dissipative nonlinear differential equations,
PNAS, 118(35) e2026805118 (2021)

\bibitem{PALMER} F. Tennie and T.N. Palmer,
Quantum Computers for Weather and Climate Prediction: The Good, the Bad and the Noisy,
arXiv:2210.17460 [quant-ph] (2022)

\bibitem{EPL_QC} S Succi, W Itani, K Sreenivasan, R Steijl,
Quantum computing for fluids: Where do we stand?
Europhysics Letters 144 (1), 10001 (2023)

\end{thebibliography}
\end{document}